\documentclass[12pt,a4]{article}\input{epsf.tex}
\begin{document}
\begin{center}
{\Large 
\bf On correctness of some processing operations for two-step cascade
 intensities data from the $(n_{th},2\gamma)$ reaction }
\end{center}
\begin{center}
{\bf A.M. Sukhovoj, V.A. Khitrov, Li Chol
}\\
{\it Joint Institute for Nuclear
Research, Dubna, Russia}\\
\end{center}

An influence of some incorrectness of analysis on the level densities and 
radiative strength functions derived from the experimental $\gamma$-spectra is 
considered. It was  shown that  the obtaining of reliable data from the
reaction $(n,2\gamma)$ requires one to derive dependence of the two-step 
cascade intensities on their primary transitions energy. The influence 
of some conditions of an analysis of the experimental $\gamma$-spectra from 
reaction $(^3He,\alpha)$ on the expected value of both level density and 
radiative strength functions was estimated. The ways to decrease these 
uncertainties are suggested.\\\\

\section{Introduction}\noindent

The form of $\gamma$-spectra measured in different experiments is ambiguously 
determined by the level density $\rho=D^{-1}$
and radiative strength functions
$k=<\Gamma_{\lambda i}>/(E_{\gamma}^3\times A^{2/3}\times
D_{\lambda})$
(or in equivalent form $f=k \times A^{2/3}$) for the $\gamma$-transitions 
between the levels $\lambda \rightarrow i$
in a nucleus with the mass $A$. The intensity of any such spectrum for 
a given $\gamma$-transition energy $E_{\gamma}$
is determined by folding of these two parameters of the $\gamma$-decay of 
a nucleus under study. Therefore, these parameters can be determined for 
any nucleus only jointly from the solution of reversed mathematical problem. 
In that case, the correspondence of the shape of functional relation 
between the measured intensity of spectrum and required mean values of 
$\rho$ and $k$ to the experiment is postulated.

In all known cases this relation includes the product  $\rho$ and $k$.
That is why, any errors of, for example, $\rho$ are completely or partially 
compensated in measured spectrum by corresponding errors of $k$. 
As a consequence, any method used for simultaneous determination of these 
parameters
 of the $\gamma$-decay by the solution of reverses problem cannot give unique 
functional dependence between $\rho$ and $k$ for given excitation and 
$\gamma$-transition energies. Moreover, all existing methods of the analysis 
allow one to determine with acceptable precision only  the sum of level 
densities of both parities in some spin window and sum of radiative strength 
functions for dipole electric and magnetic $\gamma$-transitions. 
The uncertainty in level density partially decreases if one uses additional
 experimental data on the $\rho$ below the excitation energy 1-2 MeV and 
spacing $D_{\lambda}$ between neutron resonances. The use of the experimental 
ratio $k(E1)/k(M1)$ in vicinity of the neutron binding energy $B_n$ leads to 
decrease in error of strength functions. 

Principle impossibility to determine unique quantities of $\rho$ and $k$ 
was directly fixed in method [1] used to extract these parameters from 
the dependence of the two-step cascade intensities
\begin{equation}
I_{\gamma\gamma}(E_1)=F(E_1)=\sum_{\lambda ,f}\sum_{i}\frac{\Gamma_{\lambda i}}
{\Gamma_{\lambda}}\frac{\Gamma_{if}}{\Gamma_i}=\sum_{\lambda ,f}
\frac{\Gamma_{\lambda i}}{<\Gamma_{\lambda i}> m_{\lambda i}}
n_{\lambda i}\frac{\Gamma_{if}}{<\Gamma_{if}> m_{if}},
\end{equation}
following thermal neutron capture on the energy $E_1$ of their primary transitions. 
This should be taken into account also in the analysis [2] of 
the primary $\gamma$-transition spectra of nuclear reactions. This conclusion 
follows from from testing [3] of possible use of the conventional multidimensional 
fitting of parameters of the Gauss method instead of the so-called ``Oslo method" [2].  

Multidimensional distribution of probabilities of random $\rho$ and $k$ values for 
the case of ensembles [2] of the primary $\gamma$-transition spectra, depopulating 
arbitrary excited level $E_{ex}$, is degenerate even for surplus number of experimental 
points. As a result, equal value of $\chi^2$ is observed [3] for different $\rho$ 
and $k$ even in the case when ensemble of data [2] includes experimental total 
radiative widths, densities of low-lying levels and density [4] of neutron resonances 
extrapolated into some spin interval.

For deformed nucleus, this extrapolation requires also the knowledge of dependence 
of coefficient of rotational enhancement of level density [4] on their spin. 
The lack of necessary experimental data on parameters of rotational bands even for 
the excitation energy 2-3 MeV and some higher results in unknown additional uncertainty 
in determination of $\rho$ and $k$. Of course, this question is more serious for the 
region $E_{ex}\simeq B_n$ [2].

Rather considerable non-linearity of the task, solution of which is described in [2], 
limits the variation interval of $\rho$ and $k$ even at the correlation 
coefficient $|R_{corr}|=1$ (as compared with the degenerate system of linear equations). 
Because the relation between the $\rho$ and $k$ for the primary transition spectra [2] 
and intensities of the two-step cascades [1] are essentially different, then involving
 of these heterogeneous experimental data in the united likelihood function  could 
increase reliability in determination of $\rho$ and $k$. The type and magnitude of 
the most important statistics errors at both obtaining of experimental spectra in 
the frameworks of these methods and their further use for determination of $\rho$ 
and $k$ are partially separated. This also can provide increase in reliability of the results.

Very important problem for any method for simultaneous determination of $\rho$ and 
$k$ is the minimization of additional systematic errors caused by inaccuracy in data 
processing. This follows from incompatibility of conclusions made by investigation 
[1] and [5] of two-step $\gamma$-cascades following thermal neutron capture and by 
the study of the primary $\gamma$-transition spectra of nuclear reactions [6]. 
This is reduced to the following: according to [1], cascade   $\gamma$-decay of, 
for example, deformed compound nucleus is characterized by clearly expressed 
step-like structure in level density (predicted [7] by A.V. Ignatyuk and 
Yu.V. Sokolov and related by them with breaking of the next Cooper pair of nucleons). 
But in accordance with [5,6], the main peculiarity of $\gamma$-decay is the presence of 
"pygmy" resonance in radiative strength function at "smooth" level density. 
As the authors of [1] and authors of [5,8] use the same type of experimental data 
then all observed discrepancies should be related with incorrectness of a method 
of their analysis. This incorrectness appears to a great extent due to ignoring 
[5,8] of specific of the experiment on investigation of $I_{\gamma\gamma}$.

\section{
Specific of the experiment on investigation of the two-step $\gamma$-cascades}
\noindent

Main properties of this experiment are the following:

1. the impossibility to determine ordering of two successive quanta with sum energy 
of several MeV measured with HPG detectors directly in experiment;

2. the dependence of intensity of every cascade on the $\rho$ and $k$ values over 
all excitation region of a nucleus;

3. the connection of variation $\delta I_{\gamma\gamma}$
 with variations  $\delta\rho$ and $\delta k$
 through  incoherent sum of derivatives of cascade intensities with respect to all 
possible values of $\rho$ and $k$.

   1.~As a result, every experimental spectrum 
$I_{\gamma\gamma}(E_\gamma)$
of cascades (see Fig.~1) with the total cascade energy $E_c=E_1+E_2=B_n-E_f$ 
can be reproduce by a sum 
of $2^N$ different symmetrical distributions 
$F(E_1)$ and $S(E_2)$
corresponding to registration of one or other cascade quanta as primary or
secondary transitions, respectively (here $N$ are total number of cascades).
Decomposition of the spectrum 
$I_{\gamma\gamma}(E_\gamma)$
into two parts corresponding to solely primary $F(E_1)$ and solely secondary 
$S(E_2)=F(E_c-E_1)$ transitions can be realized with some uncertainty. Practical 
determination of function $F(E_1)$ with small enough uncertainty is quite possible 
in modern experiment like 
[10] with the help of the nuclear spectroscopy methods 
using the numerical method 
[11] for improvement of energy resolution. 
Probable error of this procedure is determined by extrapolation 
[12] of cumulative 
sums of intensities of cascades (experimentally resolved in form of pair of peaks)
for small energy intervals of their intermediate levels 
$E_i<0.5B_n$ to the zero detection threshold of their intensity. 

It should be noted that any from $2^N$ possible variants of functions $F(E_1)$ and $S(E_2)$
can be always reproduced by some set of functional dependencies
$\rho(E_{ex})$ and $k(E_1)$. 
Level density for any function $F(E_1)$ is defined as the number of cascade intermediate
levels in a given interval of excitation energy independently on the truth in 
determination of their primary transition energy.
Because the value of $F$ is always determined for any 
variant of quanta ordering in cascades then in any case are also determined and 
radiative strength functions. This means that the experimental spectra of the two-step 
cascades can be reproduced, in principle, with its experimental precision
by $\rho$ and $k$ from [6] independently on the magnitude 
of systematical errors of these parameters.

The spectra for the studied even-odd and even-even nuclei demostrate 
[10] relatively 
small density of peaks for the cascades with $E_i\leq 0.5B_n$. This allows one to 
expect that the asymptotic value of error of function $I_{\gamma\gamma}(E_1)$ 
will
aspire to zero as the detection threshold of individual cascade also aspires to zero. 
The portion of unobserved cascade intensity is determined by the detection threshold 
and form of distribution of deviations of the random $\gamma$-quanta intensities 
from the average for any decaying level $E_{ex}$. If this distribution depends 
on the structure of the level $E_{ex}$ then the model notions of the 
$\gamma$-decay must be corrected both in [1] and [6].

It is very important that the specific form of dependence of $I_{\gamma\gamma}(E_1)$ on its parameters 
allows one precisely ($\chi^2/f <<1$)
reproduce this function with the help of $\rho$ and $k$ varying in narrow enough 
intervals of magnitudes. 
(Up to now, intensities of cascades $F(E_1)$ were calculated 
[1] for several millions of pairs of different functional dependencies 
of $\rho$ and $k$).

Quite different situation is observed at direct an analysis of the experimental 
spectra of the two-step $\gamma$-cascades $I_{\gamma\gamma}(E_{\gamma})=F(E_1)+S(E_2)$ whose 
amplitudes in any interval of $\gamma$-transition energy depend on both $E_1$ and $E_2$. 
Any algorithm for search of random functions (like that used in [1]) allows one 
to be convinced   that the precise reproduction of the sum of all possible values 
of functions $F$ and $S$ is possible only when the maximum values of the obtained $\rho$
 and $k$ exceed the minimum values by a factor of several tens (figs. 2,3). 
It is not difficult to check that at the fixation of $\rho$ on the ground of any 
additional information, the spectra of cascade intensities also can be exactly 
reproduced for all $E_\gamma$ by wide ensemble of $k$. Maximum value of these $k$ is 
(Fig.4,5) 2-10 times higher than minimum values (any systematic errors of 
$\rho$ shift location and distort the shape of the region of possible $k$ values).


It should be especially noted that at reproducing of intensities of two-step cascades
 in $^{172}$Yb listed in [8], "pygmy" resonance does not appear. Irremovable ambiguity
 of the experiment causes a necessity of transition from the analysis of the spectra
$I_{\gamma\gamma}(E_\gamma)$
to analysis of spectra in function of the primary transitions energy
 $I_{\gamma\gamma}(E_1)$ only.
 This circumstance is ignored by authors of both [5] and [8]. The authors of [8] make 
conclusion about correspondence of the functions $\rho^r=f(E_{ex})$ and $k^r=\phi(E_{\gamma})$ 
obtained by them from $\gamma$-ray spectra of reaction $(^3He,\alpha)$ to the experimental 
values only on the grounds that they provide reproduction of the untransformed 
experimental distributions of two-step cascade intensities in limited [5,8] interval 
of energies of their intermediate levels. The statement about correspondence between 
the experimental and tested values of $\rho$ and $k$ made on the ground of equality 
$I^{exp}=I^{cal}$ than it is equivalent to execution of conditions $F^{exp}=F^{cal}$ 
and $S^{exp}=S^{cal}$ for any energies of the cascade $\gamma$-transitions. I.e., from 
the multitude of possible solutions of eq. (2) with two unknown values of functions 
$F$ and $S$, the authors of [5,8] choose the unique solution. It is wrong mathematics 
operation. Inequality of the experimental and calculated values of $F$ and $S$ always
 will mean that the $\rho$ and $k$ parameters used for calculation of cascade intensities 
do not correspond to their unknown experimental values. Therefore, such reproduction 
($\chi^2/f<<1$) of the experimental intensities $I_{\gamma\gamma}=F(E_1)+S(E_2)$ 
 for any energies of cascade transitions is necessary but not enough criterion  of 
correspondence between the desired and unknown experimental values of $\rho$ and $k$.
Only analysis of the experimental distribution 
$I_{\gamma\gamma}(E_1)$ 
with accounting for its total error  allows determination of $\rho$ and $k$ with
 minimum possible systematic error.

   2.~The total width of the decaying initial (or intermediate) cascade level can
be presented as $\Gamma=\Gamma_{in} + \Gamma_{out}$ where two items represent sum of partial widths
 of cascade transitions in limits of energy interval chosen in [5] or [8] and out 
of it, respectively. These sums are the items of functional relation (1) of $I_{\gamma\gamma}$ with 
$\rho$ and $k$. If experimental distribution
$I_{\gamma\gamma}$ 
is reproduced with the use of some $\rho$ and $k$ only for the part of energies of 
intermediate levels and is not reproduced over the whole energy interval then, most 
probably, this partial correspondence should be considered as the result of mutual 
compensation of errors. Porter-Thomas fluctuations of the primary transition widths 
have zero mean value and cannot lead to systematic discrepancy between the 
experimental and calculated cascade intensities.

So, discrepancy between experimental and calculated cascade intensities observed 
[8] at low and high energies of cascade transitions are caused by the error in 
determination of the sum $\Gamma_{out}$ of partial widths of primary and/or secondary 
cascade transitions. In turn, this can manifest itself only at discrepancy between the 
experimental and used for calculation values of $\rho$ and $k$.

   3.~As it is known from mathematical statistics,  error of function can be 
presented in the first approach as the product of its derivative by the argument of 
function. For the eq. (1), relative errors of $\rho$ and $k$ at any energy of excitation 
or transition lead through incoherent sum of derivatives to the different relative 
change in $I_{\gamma\gamma}$ simultaneously over all energy region of cascade transitions. If this 
sum is close to zero then small relative deviation of function $I_{\gamma\gamma}$, for example, 
can be obtained only due to very wide variations of parameters or due to even small 
change in function S and necessary for this small variation of parameters of the 
calculation. In practice, in some cases 
[13] fixation of $\rho$ makes impossible 
selection of $k$ which provide reproduction of
 $I_{\gamma\gamma}(E_1)$ 
(especially at low energies of the primary transitions of cascades and large values 
of $\rho$). But there is no problem to achieve practically absolute agreement of 
the experimental and calculated sums $F+S$ for $\rho$ and $k$ with any systematical errors.

\section{
Main sources of systematic errors in different experiments}
\noindent

 Discrepancy between the main properties of the $\rho$ and $k$ values obtained in 
[1] and [5,6] means the presence of systematic error in corresponding experiments. 
The values of $\delta\rho$ and $\delta k$ in method [1] are determined, 
first of all, by the following errors of the values of the used function 
$F(E_1)$ 
as:

   (a)~the error in determination of absolute value (normalization of area);

   (b)~distortion of the form of functional dependence 
$I_{\gamma\gamma}(E_1)$ 
on the energy of the cascade primary transition.

Both errors are limited only by the experimental conditions. At the absence of 
these errors, asymptotic values of $\delta_\rho$ and $\delta_k$ for method [1] 
equals about 20\%.

Influence of the error in determination of the absolute value of 
$I_{\gamma\gamma}(E_1)$ 
on $\delta\rho$ and $\delta k$ was studied in 
[13]. Most considerable result of 
this modeling is that the sign and magnitude of errors of desired parameters 
depend on the energy of excitation and cascade $\gamma$-transition.  Besides, they 
do not depend on the error in determination of absolute values of 
$I_{\gamma\gamma}(E1)$
 at some these energies. This result confirms conclusion [1] that level density, 
for instance, cannot correspond to predictions of both model of non-interacting 
Fermi-gas 
[14] and model of constant temperature.

In practice, the minimum possible systematic errors of $\rho$ and $k$ in any 
experiment and with the use of the data processing data [1] are to be expected 
under conditions that:

   (a)~the $\gamma$-ray spectrum of the thermal neutron capture was measured within 
the method 
[15] (or within method with equivalent precision) and capture 
cross-section was determined with minimal uncertainty;

   (b)~statistics of useful events is so high that experimentally resolved 
peaks concentrate main portion of intensity of cascades with energy of intermediate 
levels $E_i\leq 0.5B_n$;

   (c)~all the other errors characterizing experiments on measurements of 
$\gamma-\gamma$ coincidences were minimized.

Because realization both of two last conditions in real experiment on measurement 
of $\gamma-\gamma$ coincidences is impossible then minimization of $\delta\rho$ 
and $\delta k$ can be achieved at careful optimization of its geometry.

Quantitative estimation of all sources of systematic errors for the data [6] 
was not performed. 

Systematic uncertainties  of determination of $\rho$ and $k$ from the spectra 
$h$ of the primary $\gamma$-transitions from the $(^3He,\alpha)$ reaction are mainly 
determined by the following:

   (a)~distortion of form of the dependence of spectrum $h$ on the primary transition 
energy at its extraction from the total experimental $\gamma$-spectra $f$;

   (b)~degeneracy of multi-dimensional distribution of random deviations of 
desired parameters from the expected value;

   (c)~necessity of model extrapolation [2] of density of neutron resonances 
into the spin region where is no experimental data.

These uncertainties can significantly change the conclusions on the cascade 
$\gamma$-decay process. So, Fig.~5 demonstrates estimation of interval of possible 
values of $k(E1)+k(M1)$ for level density used for calculation shown in Fig.~3 
(this density is less than that given in [6,8]). It maximum deviation from the 
previous variant is observed at $E_{ex}=4$ MeV and is equal to about 2/3 of 
the level density [6,8]. But even so small change in the calculated level 
density brings noticeable change in mean values of radiative strength functions 
which are necessary for reproduction of the two-step cascade intensities 
listed in [8]. For example, it changes the derivative of the strength function 
with respect to the energy. This occurs due to two factors: a necessity 
simultaneously to reproduce in calculation for increased level density of 
both cascade intensity and total radiative width $<\Gamma_\lambda>$. 
The data shown in figs.~4 and 5 point to necessity of maximum possible 
minimization of errors of the experiments under consideration.

The authors of [2,6] do not make doubts about necessity of transition from 
the experimental total $\gamma$-spectra to spectra $h$ of only primary transitions 
although this operation cannot be carried out [15] without bringing in of 
additional uncertainties in $\rho$ and $k$ even in principle. Besides, in 
some situations  the non-linearity of the error transfer (for instance, of 
the error in determination of reaction cross-section on the error of intensity 
of total $\gamma$-spectra $f$) can lead even to increase
in relative uncertainty of spectra 
$h(E_1)$ of primary $\gamma$-transitions as compared with analogous value for the
 total spectra $f_i$ of $\gamma$-rays depopulating levels $E_{ex}$. 
This conclusion follows from consideration of given in 
[16] and slightly 
modified equation
 \begin{equation}
h_1=f_1-\sum_i (h_i f_i) 
\end{equation}
Here, $i$ is the number of energy interval. The left and right parts of this 
equation must have the same  dimension (events per one decay or per
given number of decay in general case).
Therefore, all spectra $f_i$ must be normalized
to the same value as spectrum $F_1$.
If it was done with some error, for example, $f_i=\kappa_i\phi_i$, 
then it will be simply convinced that the re-normalization of the determined
from (3) spectrum $h$ to 100\% will not have error in only case when $\kappa_i=const$.
Hence, the total spectra $f_i$ must be normalized to one decay with maximum 
possible accuracy before calculation of the primary transition
spectra $h_i$  from eq. (3).
This operation must be carried out with a precision considerably exceeding the
accuracy in determination of the $\gamma$-ray intensities of the thermal neutron 
radiative capture in modern 
[15] experiment.

Estimation of permitted $\delta h$ for the distorted spectra of any kind can 
be easily obtained in the following way: the total $\gamma$-ray spectrum $f$ for
 any energy interval is recurrently calculated for the arbitrary $\rho$ and 
$k$ without statistic errors beginning from the known $\gamma$-transitions 
depopulating decay of low-lying discrete levels according to the relation
   \begin{equation}
f_1=h_1+\sum_i (h_i f_i). 
\end{equation} 
Then, the spectra of primary transitions are restored from the spectra 
distorted by random error like  $f_i=(1+e_i) f_i$ according to eq. (3).
The worst variant is the case when the error $e_i$ is proportional to the number 
of spectrum and increases its intensity as increasing $i$ 
(for numeration used in 
[16]). 

Using this technique one can obtain that maximum relative error of 
normalization of spectra to one decay must be considerably less than 1\%. 
This follows from the fact that at increasing of $E_{ex}$ small value of $h$ 
for low energy $E_{\gamma}$ is determined from difference of two much larger 
intensities of the total $\gamma$-spectra with very intence
low energy secondary transitions.
 Moreover, the same energy of 
$\gamma$-quantum in spectrum $f$ can be related with the sum of intensities 
of the first, second, third and so on cascade quanta at increasing of 
their intensities when excitation energy decreases.

One can try to achieve appropriate accuracy using normalization of 
total $\gamma$-ray spectrum to one decay with the help of condition 
$\sum (E_{\gamma} \times I_{\gamma})=E_{ex}$.
 Such normalization excludes the problems 
[16] of different multiplicity 
of quanta in total $\gamma$-ray spectra for different excitation energy of 
decaying levels and for precise determination of reaction cross-section. 
But this can be done, probably, only partially. Detailed investigation
of influence of energy resolution of detectors and functions of their 
response on systematic error can be required in this case, as well. 
The open question is systematic error connected with widening of spin 
window of levels excited by cascade transitions.

Statistic error of the $h$ data given in [2] is larger than 1\%. 
Therefore, it is worth while to make extraction of the primary 
$\gamma$-transition spectrum for minimization of systematic errors by means
of minimization of sums
$\chi^2=\sum(f_1^{exp}-(h_1+\sum_i (h_i f_i^{exp}))^2$
instead of mathematical method used in procedure 
[16].

\section{
Systematic errors of joint analysis of different experiments and possibilities
 of their decreasing}
\noindent

Level densities and radiative strength functions presented in [6], 
in opinion of authors of this work, correspond to their most probable 
values because uncertainties of method [2] were completely eliminated 
due to fixation of level density at two excitation energies and the 
use of the experimental total radiative width of neutron s-resonances.
Correctness of this statement depends
only on the form of multi-dimensional distribution of probability of random 
values of desired parameters.

As it was obtained in [3], deviation distribution of the $\rho$ and $k$ values 
from their averages for the experimental data [6] is generated owing to 
correlation of these parameters (even at at the use in [6] of complementary 
with respect to [2] information). One can try to decrease this degeneracy or 
even eliminate it only by involving complementary information in the maximum 
likelihood function.

Effective search for its maximum and identification of false solutions is 
realized now by means of the Gaussian method for solving systems of nonlinear 
equations. At any their degeneracy, the region of possible solutions is found 
within the programs of multi-parameter fitting with the use of the procedure 
of regularization of covariant singular matrix with obligatory variation of 
input values of desired parameters. The presence of exponentially changing 
parameters leads to appearance [3] of non-principle problems for the use of 
modern programs which realized the Gaussian method for determination of the most 
probable $\rho$ and $k$ values for the set of the data [6]. As the experience 
[3] shown, there is no grounds to wait principle obstacles  for involving of 
derivatives of function $I_{\gamma\gamma}$
from eq.~(1) with respect to its parameters in the Jacoby matrix.

Unfortunately, there is no any grounds to hope for complete exclusion of 
systematic errors of $\rho$ and $k$ even in this variant of analysis. 
Any hypotheses about the cascade $\gamma$-decay process of an excited nucleus 
(directly or indirectly used in [1], [6,8]) also lead to appearance of 
additional systematic errors of the desired parameters of the cascade $\gamma$-decay 
process. But the use of them is absolutely necessary, for instance, above 
the excitation energy 1-3 MeV of deformed nucleus due to the lack of any 
required experimental information in this region. 

First of all this concerns the problem of the shape of the energy dependence 
of the radiative strength functions of equal multipolarity and $\gamma$-quantum 
energy for different energies of decaying levels. Available data (theoretical 
calculation of matrix elements, experimental data [4] and intensities of the 
cascade population of levels up to the excitation energy $\leq 1$ MeV
 [17] and higher 
[18]) show that influence of this factor on systematic error 
cannot be neglected.

This problem common for both methods [1] and [2]. But there is and 
considerable difference: in the case of the data [6], resulting value of $k$ 
is a superposition  of strength functions for $\gamma$-transitions of equal 
energy but depopulating levels with sufficiently different energy and 
structure of wave functions. In the case [1], difference between the energy 
dependences of strength functions of the secondary transitions and analogous 
dependence of the primary transitions affects the desired parameters $\rho$ 
and $k$ only indirectly - through distortion of the $\Gamma_{if}$ value. 
At low energy of the secondary transitions this distortion decreases due to 
both factor $E_\gamma^3$ and possible [18] different in sign change in the shape 
of dependence $k=\phi(E_\gamma)$ for different energy of $\gamma$-transitions 
(for equal energy of decaying level).
\section{Conclusion}
\noindent

Insufficient degree of correctness of using mathematics and mathematics 
statistics for extraction of the level density and radiative strength 
functions from the primary $\gamma$-transition spectra of nuclear reactions 
and spectra of the two-step $\gamma$-cascades leads to mistaken conclusions 
about the process under study. Therefore, the use of the analysis procedure
suggested in 
[19] cannot provide obtaining of the reliable data on the 
radiative strength functions.\\\\

{\large\bf References}

\begin{flushleft}

1. E.V. Vasilieva, A.M.Sukhovoj, V.A. Khitrov,
Phys. At. Nucl., {\bf 64(2)} (2001) 153.\\
\hspace*{14pt}  V.A. Khitrov, A.M. Sukhovoj,
 INDC(CCP)-435, Vienna, 21 (2002).
\\\hspace*{14pt}
http://arXiv.org/abs/nucl-ex/0110017\\
2. A. Schiller et al., Nucl. Instrum. Methods Phys. Res.,  {\bf A447} (2000) 498.\\
3. V.A. Khitrov at al.,
In:  XI International Seminar on Interaction
of Neutrons
\\\hspace*{14pt}with Nuclei  (Dubna, May 2003) Dubna, 2004, E3-2004-22, p.107 \\
4. "Handbook for calculation of nuclear reaction.
data", IAEA-TECDOC-1034, 1998, \\
5. F. Be\v{c}v\'{a}\v{r} \sl et al.\rm, Phys.\ Rev.\ C \bf 52\rm,
1278 (1995).\\
6. A.Voinov,  M. Guttormsen, E. Melby, J. Rekstad, A. Schiller,
 S. Siem, \\
\hspace*{14pt}Phys.\ Rev.,\ C \bf 63(4)\rm,044313-1 (2001).\\
7. A.V. Ignatyuk,  Yu.V. Sokolov
Yad. Fiz.,  {\bf 19} (1974) 1229.\\
8. A. Schiller \sl et al.\rm, preprint nucl-ex/041038.\\
9. J.Honz\'atko et al., Fizika B (Zagreb) {\bf 12(4)} (2003) 299.\\
10. V.A. Bondarenko et al., Fizika B (Zagreb) {\bf 11} (2002) 201.\\
11. A.M. Sukhovoj, V.A. Khitrov, Instrum. Exp. Tech., {\bf 27} (1984) 1071\\
12. A.M. Sukhovoj, V.A. Khitrov, Phys. of Atomic Nuclei, {\bf 62(1)} (1999) 19\\
13. V.A.Khitrov,  Li Chol, A.M.Sukhovoj, 
In: XI International Seminar on Interaction\\\hspace*{14pt}
of Neutrons with Nuclei,  Dubna, May 2003,
E3-2004-22, Dubna, 2004, p. 98.\\
14. W. Dilg, W. Schantl, H. Vonach and M. Uhl, Nucl. Phys., {\bf A217} (1973)
 269.\\
15.  G.L. Molnar et al., App. Rad. Isot., {\bf 53} (2000) 527\\
\hspace*{14pt}
http://www-nds.iaea.org/pgaa/egaf.html\\
16. M. Guttormsen at all, Nucl.\ Instrum.\
Methods Phys.\ Res.\ A \bf 255\rm (1987) 518.\\
17. A.M.Sukhovoj, V.A.Khitrov,
Physics of Atomoc Nuclei, {\bf 67(4)} (2004) 662.\\
18. V.A. Bondarenko et all, In: XII International Seminar on Interaction\\\hspace*{14pt}
of Neutrons with Nuclei,  Dubna, May 2004,
E3-2004-50, Dubna, 2004, p. 18.\\
19. A. Voinov \sl et al.\rm, Nucl.\ Instrum.\ Methods Phys.\ Res.\ A
\bf 497\rm, (2003) 350.\\
\end{flushleft}

\begin{figure}
\vspace{2cm}
\begin{center}
\leavevmode
\hspace{-.8cm}

\epsfxsize=16cm

\epsfbox{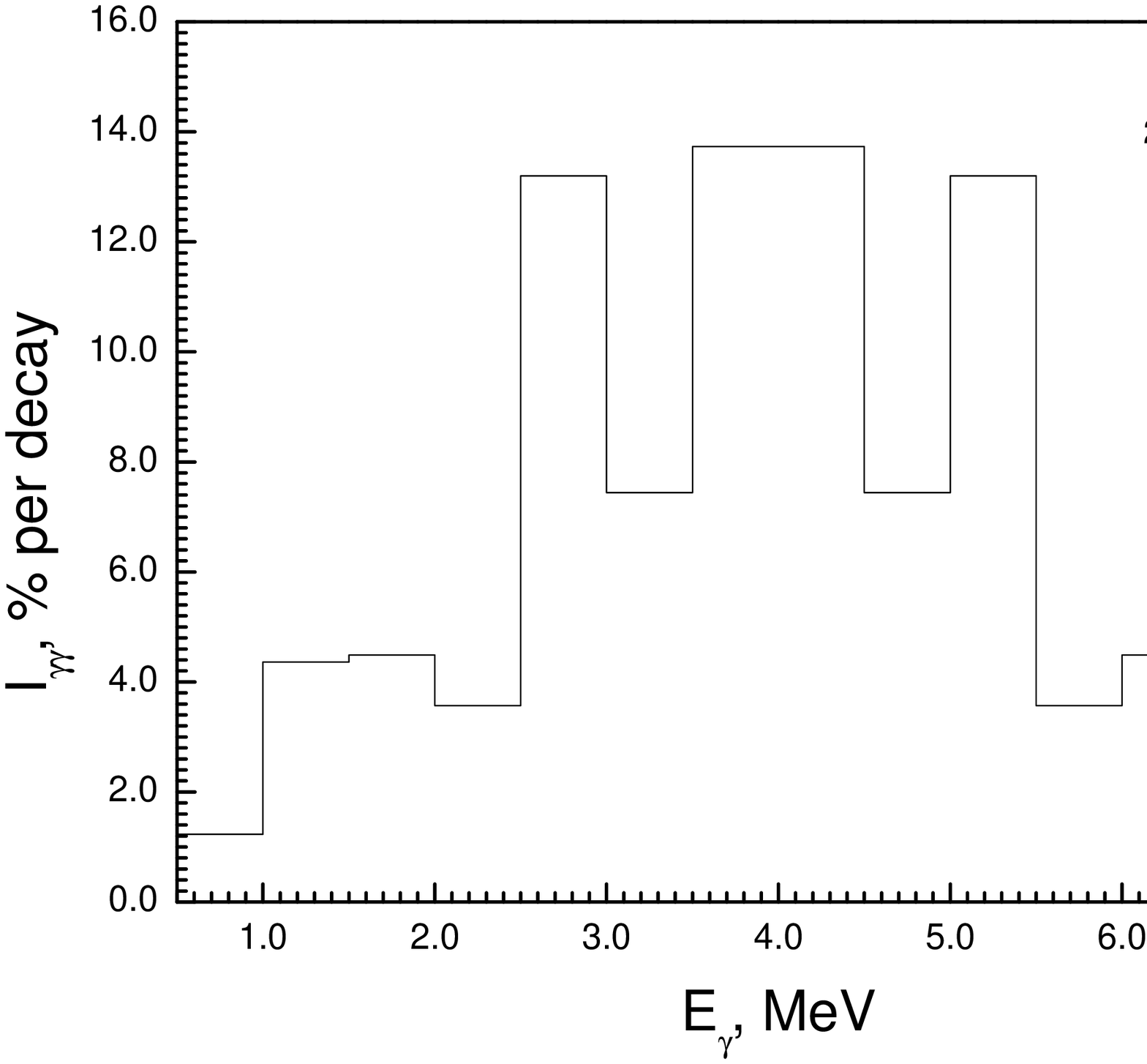}

\epsfxsize=16cm

\epsfbox{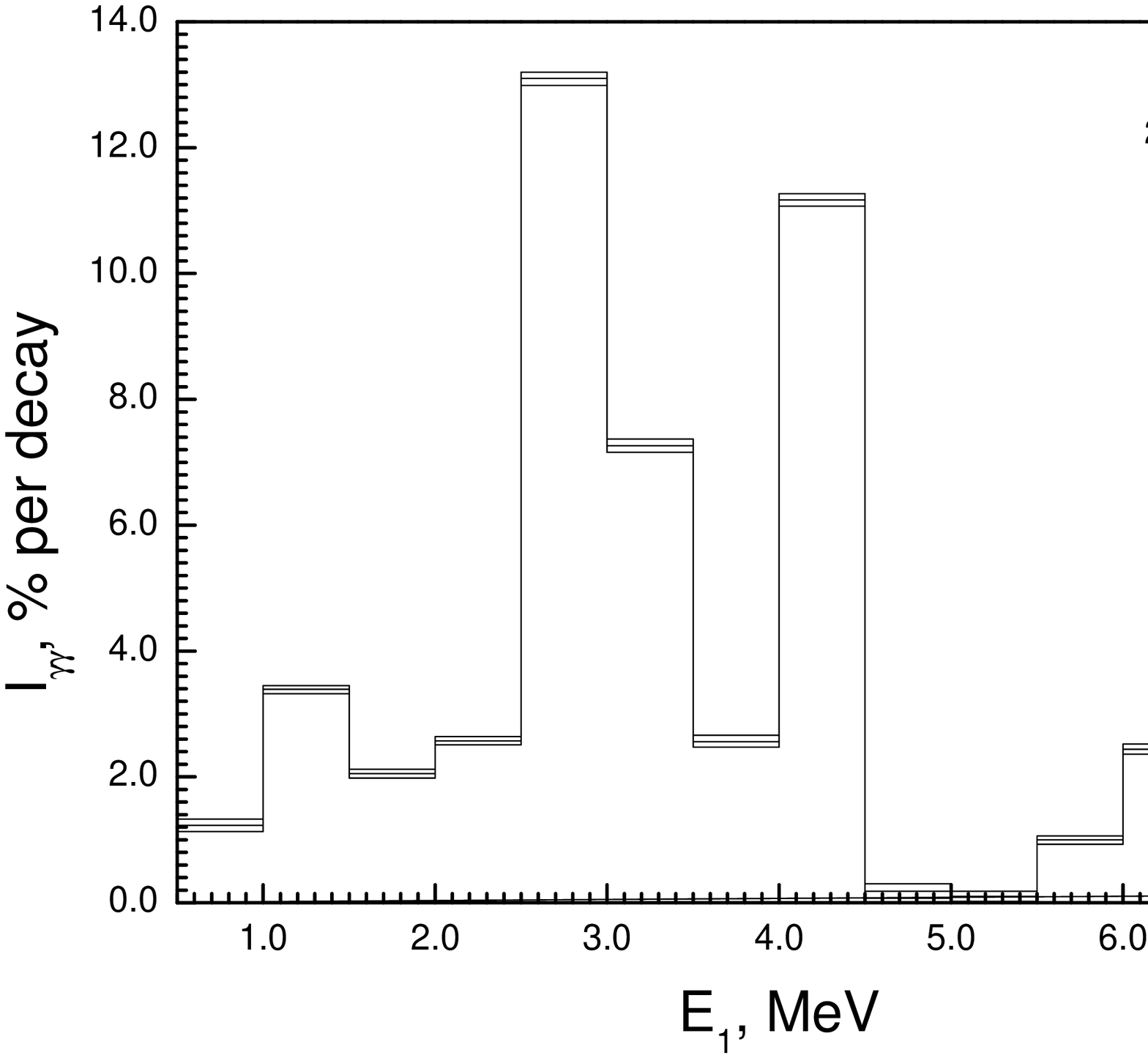}
\end{center}
\vspace{-5cm}

Fig.~1.~Top: the experimental
spectrum of the two-step $\gamma$-cascades terminating at the first and 
ground states of $^{28}Al$. Bottom: 
~distribution [9] of the same intensity 
in function of the primary
transition energy only.
\end{figure}
\newpage

\begin{figure}
\vspace{2cm}
\begin{center}
\leavevmode
\hspace{-.8cm}

\epsfxsize=16cm

\epsfbox{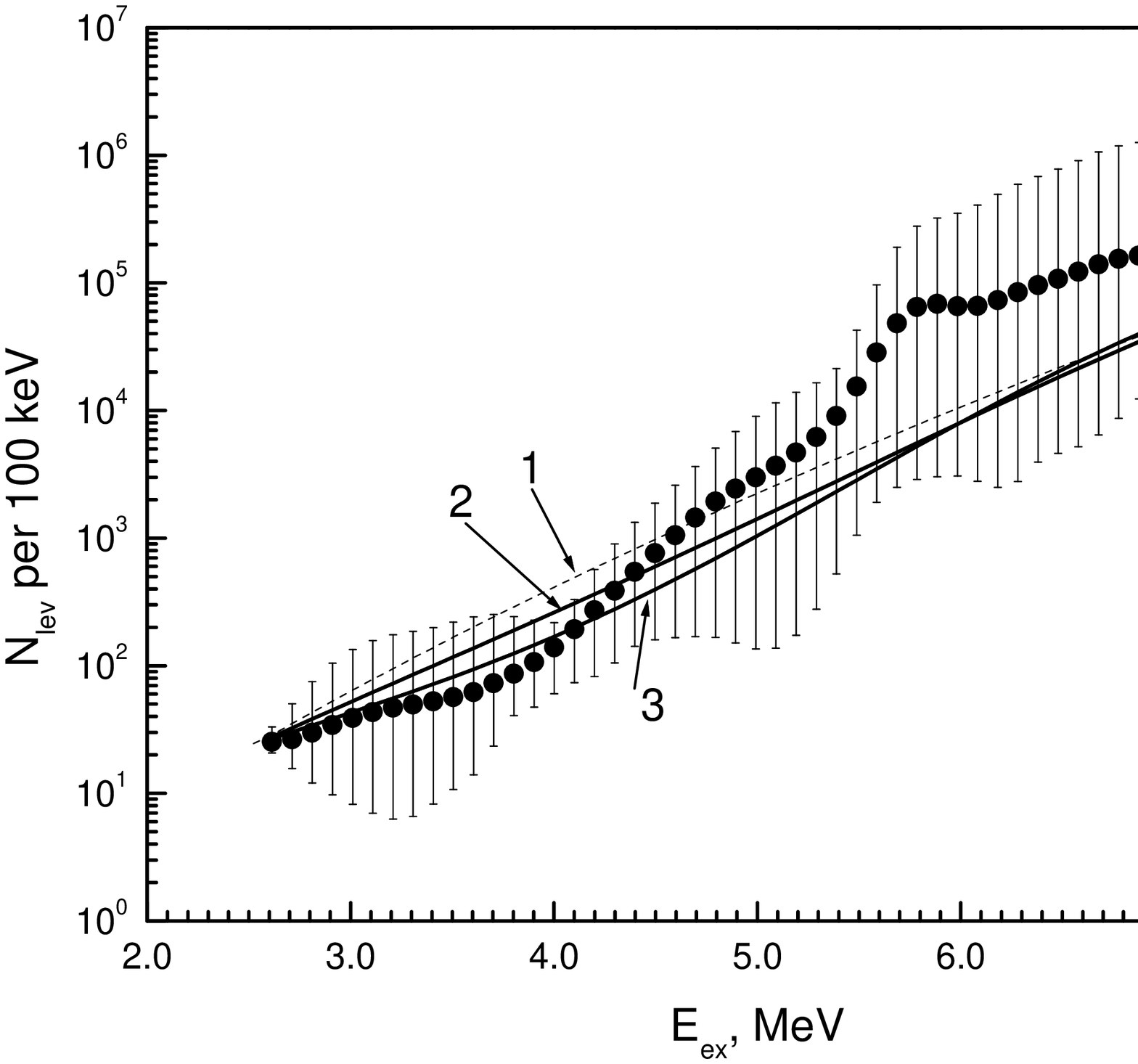}
\end{center}
\vspace{-5cm}
Fig.~2.~The interval of variations of the level density providing reproduction
 of the experimental intensities [8] of the two-step cascades to the ground and 
first excited state of $^{172}Yb$ (points with bars). Line 1 represent 
predictions according model [14], lines 2 and 3 show the data used in 
calculation for figs. 4 and 5, respectively.
\end{figure}
\begin{figure}[htbp]
\vspace{3cm}
\begin{center}
\leavevmode
\epsfxsize=16cm

\epsfbox{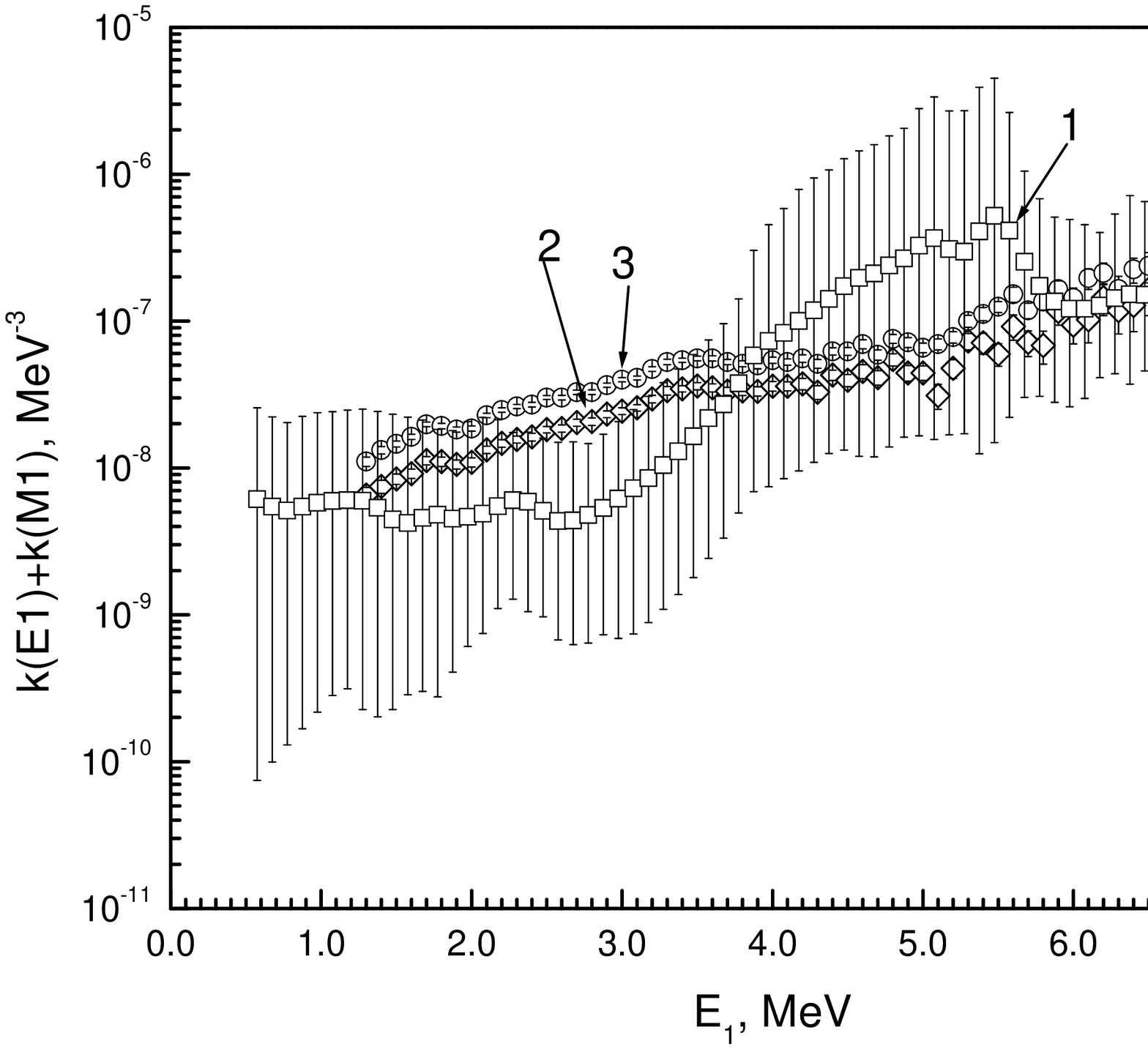}
\end{center}
\vspace{-5cm}
Fig.~3.~Interval of variations of the strength functions (points 1 with bars).
Points 2 and 3 represent strength functions from [6] and [8], respectively.
\end{figure}
\newpage
\begin{figure}[htbp]
\vspace{1cm}
\begin{center}
\leavevmode
\epsfxsize=16cm

\epsfbox{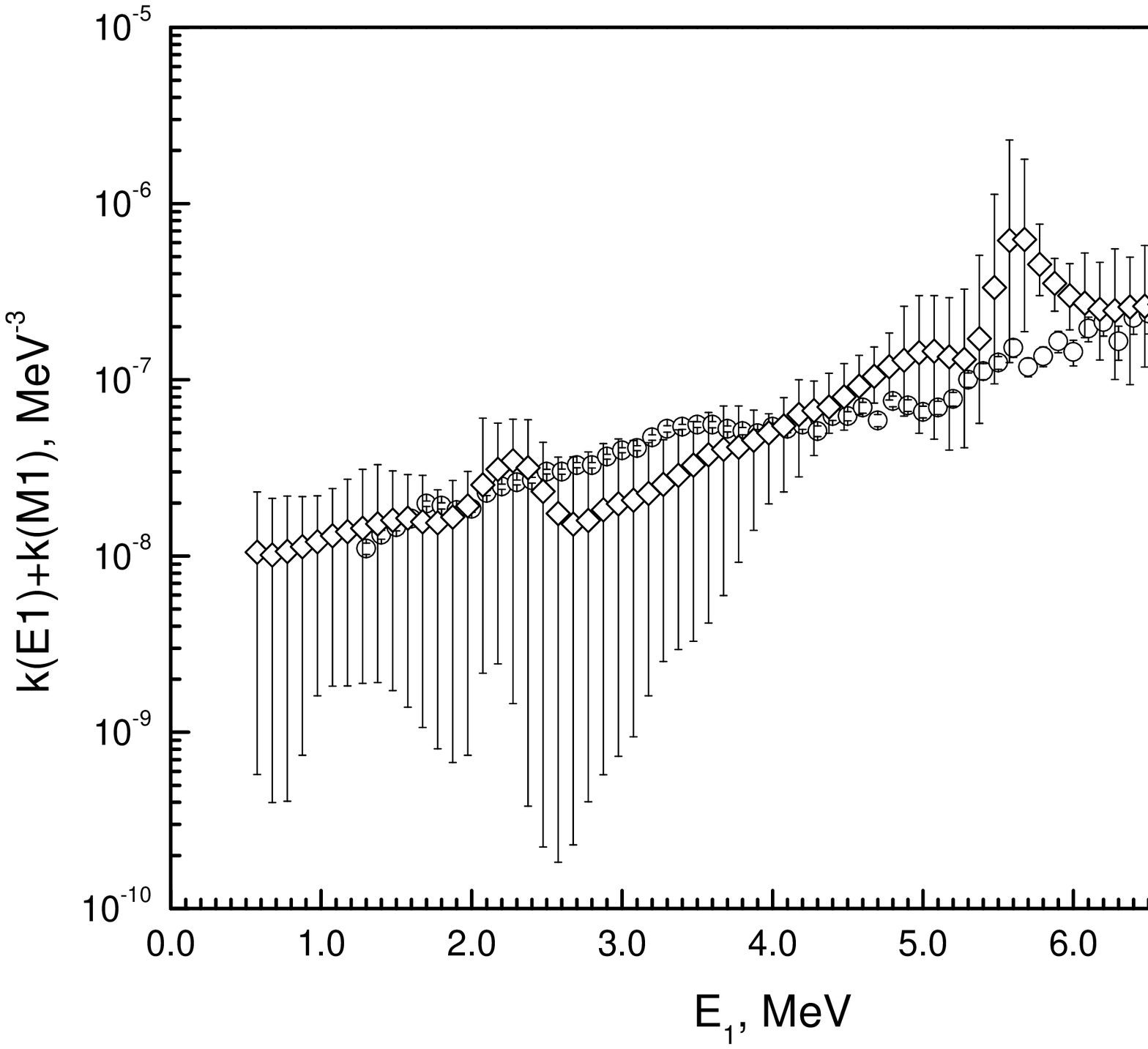}
\end{center}
\vspace{-5cm}
Fig.~4.~The same as in Fig.~3. 
Calculation with level density shown in Fig.~2 by line 2.
\end{figure}
\begin{figure}[htbp]
\vspace{1cm}
\begin{center}
\leavevmode
\epsfxsize=16cm

\epsfbox{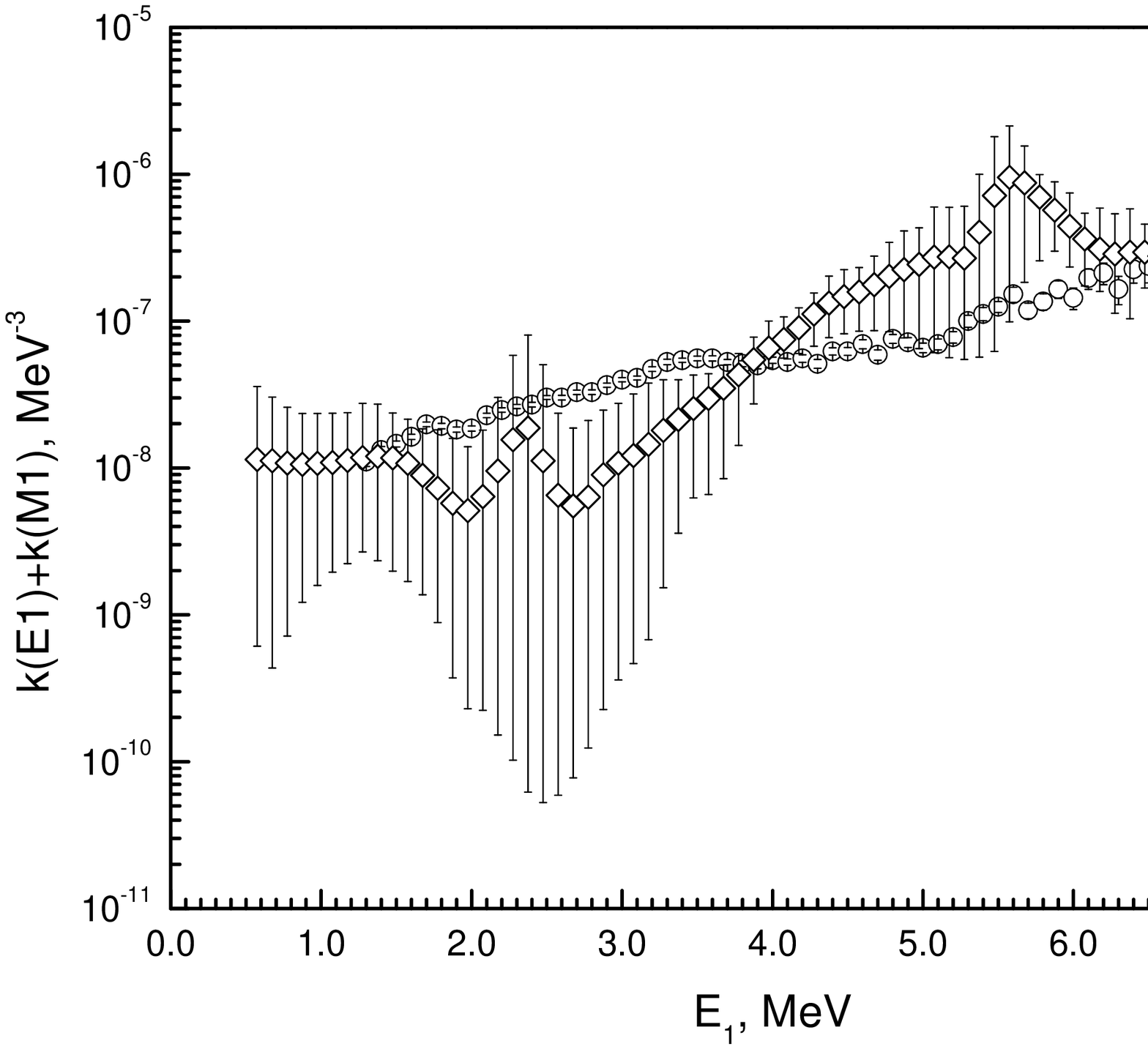}
\end{center}
\vspace{-5cm}
Fig.~5.~The same as in Fig.~3. 
Calculation with level density shown in Fig.~2 by line 3.
\end{figure}
\end{document}